\documentclass[manuscript]{acmart}
\setcopyright{rightsretained}
\acmConference{AutomationXP26 Workshop of the 2026 CHI Conference on Human Factors in Computing Systems, April 14, 2026, Barcelona, Spain}{}{}
\acmDOI{}
\acmISBN{}
\usepackage{booktabs}
\usepackage[table]{xcolor}
\usepackage{float}
\usepackage{enumitem}

\title{From Role to Person: Trust Calibration Challenges in Twin Agents}

\author{Hugo Andersson}
\affiliation{
  \institution{Aarhus University}
  \city{Aarhus}
  \country{Denmark}
}
\email{hugo@cs.au.dk}

\author{Niklas Elmqvist}
\affiliation{
  \institution{Aarhus University}
  \city{Aarhus}
  \country{Denmark}
}
\email{elm@cs.au.dk}

\begin{document}

\begin{abstract}
    Agentic AI has taken on the role of assistant, collaborator, and decision-support tool.
    We argue the next role on that list is more personal: you.
    These are digital twins of each individual---\textit{twin agents}---representing their knowledge, perspective, and communicative style to colleagues when they are unavailable.
    Drawing on early design work in an ongoing project in which agents represent knowledge workers in a professional setting, we identify a trust calibration problem specific to this approach.
    When a human colleague doubts a twin agent's output, they face three failure modes (a schema gap, an epistemic gap, and a model artifact) with no reliable attribution path between them.
    Cognitive forcing functions and related frameworks address overreliance effectively in contexts where there is a clear boundary between the AI and the human decision-maker.
    However, twin agents dissolve that boundary, raising a class of trust calibration challenge these frameworks were not designed to handle.
    We introduce the concept, distinguish it from digital twins, and outline the research questions this new class of agent demands.
\end{abstract}


\maketitle
\pagestyle{plain}
\thispagestyle{plain}

\section{Introduction}

As agentic AI systems grow more capable, their trajectory is moving past tasks and assistant roles toward something more personal: agents that stand in for a real human being in a professional setting.
Recent work already demonstrates systems that act on a person's behalf in real-time communication~\cite{cheng2025conversational}, and AI agents that replicate managerial personas have recently been examined in the research literature~\cite{hu2026boss}.
More foundational work moves toward agents that carry persistent knowledge about specific individuals, from generative agents that simulate social behavior~\cite{park2023generative} to user models built from behavioral traces~\cite{shaikh2025creatinggeneralusermodels}.
We see the next step in this trajectory as an emergent consequence of these developments, not an agenda we advocate: agents that no longer merely assist a person, but stand in for them.

We call these \textbf{twin agents}.
In this paper, we introduce the concept, distinguish it from related work, and identify a trust calibration problem that arises specifically in twin agent contexts and that existing frameworks are not equipped to handle.
We approach this from what Ehsan and Riedl~\cite{ehsan2020hcxai} call a \textit{reflective sociotechnical} stance: rather than asking how to make twin agents more capable or accurate, we ask what this approach does to the humans navigating it; the colleagues who must decide what to believe, and the people whose identities are being represented.
What we call twin agents have already appeared in early organizational deployments~\cite{gani2025klarna, hu2026boss}. The technical foundations for person-specific simulation are advancing rapidly~\cite{park2024generative, simile2026}, and we see these together as early indicators of a trajectory worth examining now.

\section{Twin Agents}

A \textit{twin agent} is a social AI agent grounded in the communicative and epistemic profile of a specific real individual, constructed to represent rather than replace them.
It is a professional representation, reflecting how someone thinks, communicates, and engages with colleagues in a work context: their expertise, their perspective, and their characteristic ways of reasoning.
A twin agent is a social and epistemic stand-in that is active in contexts where the real person is absent.
It should \textbf{never} impersonate that person or commit actions on their behalf.

The term is derived from \textit{digital twins}, a well-established concept in engineering referring to computational replicas of physical systems used for simulation and monitoring.
The comparison is instructive precisely because the two concepts diverge on every dimension that matters for organizational AI (Table~\ref{tab:comparison}). 
Where digital twins strive for maximum replication fidelity, twin agents treat impersonation as a \textbf{failure mode}.
Where digital twins are passive models queried by other systems, twin agents are active social actors interacting with people who have real relationships with the person being represented.
The ``twin'' framing is chosen deliberately: twins share origin and close resemblance, but they are clearly \textbf{not} the same entity.
In other words, a twin agent is not a copy or a clone; it is a related but separate presence, and the gap between them is where the trust problem lives.

At this point, it might be worthwhile questioning why twin agents should even be considered given all of the ethical concerns and potential pitfalls. However, the motivation behind twin agents is straightforward: expertise is increasingly decoupled from physical presence, but it remains coupled to availability.  While distributed work has normalized asynchronous collaboration, the knowledge, judgment, and perspective of a specific person still require that person to be reachable. Twin agents are a response to that constraint. Recent work has demonstrated that agents trained on interviews with real individuals can replicate their attitudes and behaviors with measurable accuracy~\cite{park2024generative}, and the commercial translation of that research is already underway~\cite{simile2026}. The potential is significant: twin agents could fundamentally change how expertise travels through organizations, making knowledge and judgment available in ways that current tools cannot approximate. Twin agents are, in that sense, an emerging reality that demands HCI attention before the design space closes around assumptions we have not yet examined.

We note that twin agents also raise important questions about the agency, consent, and identity rights of the person being modeled.
What it means for a professional identity to be represented, approximated, and potentially distorted without one's continuous involvement is a distinct and equally pressing problem.
We focus here on the interaction-level trust challenge faced by the colleague on the receiving end, and treat the represented person's perspective as a companion research question.

\begin{table}[h]
\caption{\textbf{Comparison.} Digital Twins vs.\ Twin Agents.}
\label{tab:comparison}
\renewcommand{\arraystretch}{1.4}
\rowcolors{2}{gray!12}{white}
\begin{tabular}{p{2.8cm}p{3.8cm}p{6.2cm}}
\toprule
 & \textbf{Digital Twins} & \textbf{Twin Agents} \\
\midrule
Origin domain & Engineering, manufacturing & Organizational AI, HCI \\
What is modeled & Physical or biological systems & Social, epistemic, communicative identity \\
Fidelity ideal & Maximum accuracy; more is always better & Representation, not replication; impersonation is a failure mode \\
Mode & Passive model queried by systems & Active social actor interacting with people \\
Who interacts & Other systems, engineers & Colleagues with real relationships to the represented person \\
Primary use & Simulation, monitoring, prediction & Coordination, knowledge relay, asynchronous presence \\
\bottomrule
\end{tabular}
\end{table}

\section{Related Work}

Overreliance on AI is a well-documented problem: even when explanations and confidence scores are provided, people tend to accept AI recommendations uncritically~\cite{bucinca2021trust}. Cognitive forcing functions, interventions requiring deliberation before accepting a recommendation, can help, but only when a legible boundary exists between the AI and the human decision-maker.

On the technical side, Park et al.~\cite{park2023generative} demonstrate that LLM agents can sustain contextually consistent, socially believable 
behavior over extended interactions in simulated environments, establishing the plausibility of agents as persistent social actors, though in their 
work the agents represent fictional characters rather than specific real individuals. Building on this, Park et al.~\cite{park2024generative} demonstrate that agents trained on qualitative interviews with real people can replicate their attitudes and behaviors with measurable accuracy, establishing the feasibility of person-specific simulation at scale. Shaikh et al.~\cite{shaikh2025creatinggeneralusermodels} close this gap by introducing GUM, a system that constructs rich, confidence-weighted models of specific individuals from passive observation 
of behavioral traces such as screen recordings and communications. GUM demonstrates that detailed person-models can be inferred without explicit 
elicitation, positioning passive inference as a scalable path toward individual-level agent representation.

Recent work has also begun exploring agents that represent specific real 
individuals in organizational contexts. Cheng et al.~\cite{cheng2025conversational} examine conversational agents that communicate on a person's behalf during multitasking, surfacing questions of shared autonomy and appropriate delegation. Hu et al.~\cite{hu2026boss} push this further by exploring AI systems that replicate managerial personas, finding through design workshops that higher fidelity may unsettle rather than reassure.

Twin agents sit at the intersection of these threads: like GUM, they are grounded in a specific real person; like generative agents, they are active social actors. Unlike both, they introduce a new class of trust and attribution problem rooted in their social role within real organizational relationships. This problem has a foundation in both social and psychological research: the CASA (Computers Are Social Actors) tradition established by Nass and Moon~\cite{nass2000machines} shows that people automatically apply social heuristics to machines regardless 
of awareness. Furthermore, recent work on AI-generated language demonstrates that people cannot reliably detect it, with well-calibrated systems producing text perceived as more natural than human-written output~\cite{jakesch2023human}.

\section{The Attribution Problem}

This attribution problem surfaced during the design and development of a simulated workspace in which agents representing real knowledge workers inhabit a shared environment.
We designed the agent pipeline and before building, reasoned through the ethical and sociotechnical implications of deploying agents that represent real colleagues. These were engineering and design conversations, not empirical findings, but they revealed a structural problem that warrants attention. The threefold ambiguity below is, in that sense, an early warning from a system still under construction, one we intend to pursue empirically as our framework develops and its agents begin operating in realistic professional settings.

When a colleague's twin agent tells you something, you are not receiving a system recommendation. You are receiving what is framed as your colleague's knowledge and perspective. The trust you extend is not trust in a model; it is trust in a person, mediated by a model. This reframing creates a new attribution problem. When a twin agent's output causes a human colleague to doubt it, that colleague faces three simultaneous explanations that cannot be disentangled from the outside:

\vspace{8pt}

\begin{itemize}[topsep=2pt, itemsep=1pt, parsep=0pt]
  \item \textbf{Schema gap}: the agent's representation of the person is incomplete or inaccurate
  \item \textbf{Epistemic gap}: the user simply does not know the person's actual view on this matter
  \item \textbf{Model artifact}: the output reflects an LLM failure such as hallucination, sycophancy, or drift
\end{itemize}

\vspace{8pt}

The schema gap and epistemic gap may feel identical to the doubting colleague --- both surface as uncertainty about what the person thinks --- but they originate in different places: one in the agent's model of the person, the other in the colleague's own knowledge of them. In prior work on overreliance, there is a ground truth: the AI was wrong, the user followed it anyway. In the twin agent case, ``wrong'' is itself ambiguous. Was the agent wrong about what the colleague thinks, or was the user wrong about what the colleague thinks? No reliable attribution path exists from the outside, and this ambiguity has downstream consequences for trust, accountability, and team dynamics.

This has structural consequences for how trust can be calibrated over time. Bansal et al.~\cite{bansal2019beyond} show that effective human-AI teaming depends on humans forming an accurate mental model of the AI's \textit{error boundary}: the conditions under which it fails. That model is learnable, but only when ground truth feedback is available. Twin agents break this assumption entirely. The relevant ``error'' is not a task failure with a verifiable outcome but a misrepresentation of a person, and colleagues have no independent access to the real person's views to calibrate against. Every interaction with the agent is simultaneously data about the agent \textit{and} data about you. The error boundary is structurally unlearnable, which is precisely why twin agents require a different design logic than task-executing agents.

\begin{figure}[H]
  \centering
  \includegraphics[width=\columnwidth]{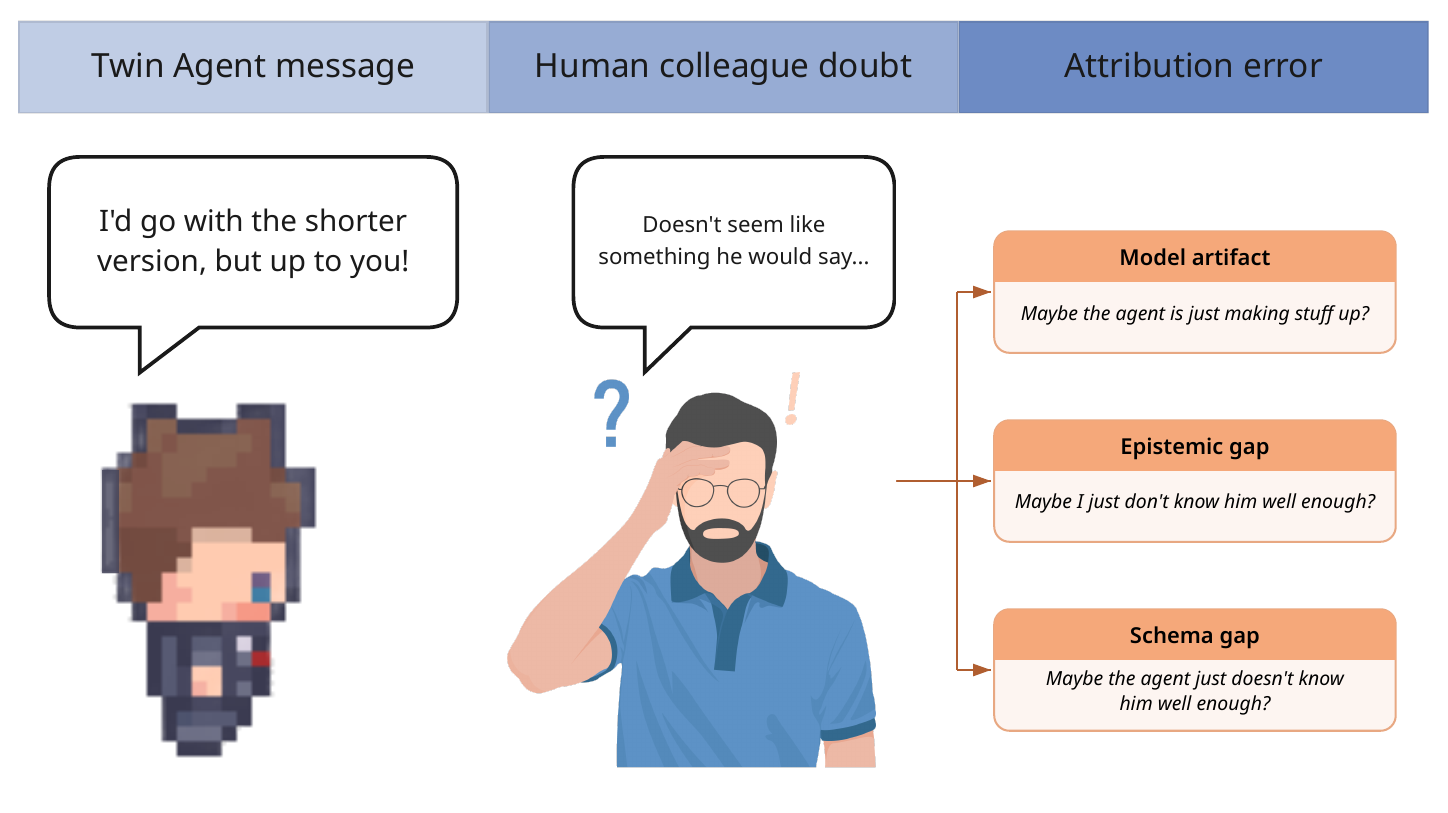}
  \caption{\textbf{Agent behavior.} A twin agent (left, represented as an avatar) sends a message to a human colleague, who reacts with doubt. The colleague cannot determine the source of that doubt: it may stem from a \textit{schema gap} (the agent's representation of the person is incomplete), an \textit{epistemic gap} (the colleague simply does not know the person's view), or a \textit{model artifact} (an LLM failure such as hallucination, sycophancy, or drift). All three explanations are consistent with the same observed output, leaving the colleague with no reliable attribution path.}
  \label{fig:attribution}
\end{figure}

\section{Research Agenda}

Twin agents open a rich design space with numerous compelling use cases. Beyond coordination, twin agents could enable entirely new workflows: simulating how a colleague might respond to a proposal before presenting it, generating research collaboration ideas by combining two people's epistemic profiles, or producing drafts grounded in someone's known positions and reasoning style. The same properties that create the attribution problem we describe — the blurring of agent and person, the relational stakes, the epistemic depth required — also create opportunities for richer, more human forms of asynchronous collaboration than current tools afford. We see three directions for future work that explore this space, while addressing the trust challenges it introduces.

First, interface design needs to distinguish between \textit{layers of agent output} rather than merely signal agent involvement. Knowing a message came from a twin agent is insufficient for calibration if the recipient cannot tell whether the agent is synthesizing an inferred view or relaying something the person actually authored. A finer transparency layer that distinguishes reconstructed inference from direct relay would give colleagues a more actionable basis for trust than a generic disclosure badge.

Second, attribution legibility may be partially recoverable through \textit{epistemic provenance}. When a twin agent's output is traceable to a specific artifact authored by the represented person --- a design document, a stated position, a prior communication --- surfacing that source transforms the schema gap from invisible to inspectable. The colleague can evaluate the agent's claim against something the person actually produced, rather than against an opaque model. A twin agent that can say ``this reflects what she wrote in the project brief'' is meaningfully different from one that cannot, and that difference is a design resource, not merely a logging concern.

Third, the relational dimension of twin agent trust requires a new class of intervention that cognitive forcing functions were not designed for. Existing overreliance interventions assume a socially neutral relationship between a user and a system. Twin agents dissolve that assumption: the colleague is not evaluating a model; they are navigating a pre-existing relationship, and the social cost of expressing doubt is itself a variable. An intervention that would work between a user and an anonymous AI recommendation system may be ineffective --- or actively counterproductive --- when doubt implicates a real colleague. Interventions for twin agents need to operate at the relational level: making visible when an agent is operating beyond the range of its documented representation, and creating low-cost pathways for colleagues to surface unresolved uncertainty back to the real person. The goal is a form of calibrated relational trust that accounts for epistemic uncertainty while preserving the social value that makes twin agents worth deploying in the first place.

\section{Conclusion}

Twin agents offer utility by relaxing the constraints of time and space that tie a person's expertise to their physical presence. But the trust problem they introduce is not a refinement of existing challenges --- it is a different kind of problem. When the AI stands in for a colleague, trust calibration can no longer rely on a legible boundary between the two. The threefold attribution ambiguity we describe is not an engineering problem; it is a structural feature of what twin agents are. We suggest the HCI community develop theories and interventions grounded in the relational dimension of human-AI trust, not just its epistemic one.

\bibliographystyle{ACM-Reference-Format}
\bibliography{references}

@inproceedings{ehsan2020hcxai,
  author    = {Ehsan, Upol and Riedl, Mark O.},
  title     = {Human-centered Explainable {AI}: Towards a Reflective Sociotechnical Approach},
  booktitle = {Proceedings of the AAAI Workshop on Artificial Intelligence Safety (SafeAI)},
  year      = {2020},
  publisher = {AAAI Press},
  note      = {arXiv:2002.01092}
}

@article{nass2000machines,
  author    = {Nass, Clifford and Moon, Youngme},
  title     = {Machines and Mindlessness: Social Responses to Computers},
  journal   = {Journal of Social Issues},
  volume    = {56},
  number    = {1},
  pages     = {81--103},
  year      = {2000},
  publisher = {Wiley},
  doi       = {10.1111/0022-4537.00153}
}

@inproceedings{bansal2019beyond,
  author    = {Bansal, Gagan and Nushi, Besmira and Kamar, Ece and Lasecki, Walter S. and Weld, Daniel S. and Horvitz, Eric},
  title     = {Beyond Accuracy: The Role of Mental Models in Human-{AI} Team Performance},
  booktitle = {Proceedings of the Seventh AAAI Conference on Human Computation and Crowdsourcing},
  year      = {2019},
  publisher = {AAAI Press},
  doi       = {10.1609/hcomp.v7i1.5285}
}

@article{jakesch2023human,
  author    = {Jakesch, Maurice and Hancock, Jeffrey T. and Naaman, Mor},
  title     = {Human Heuristics for {AI}-Generated Language Are Flawed},
  journal   = {Proceedings of the National Academy of Sciences},
  volume    = {120},
  number    = {11},
  pages     = {e2208839120},
  year      = {2023},
  publisher = {National Academy of Sciences},
  doi       = {10.1073/pnas.2208839120}
}

@inproceedings{bucinca2021trust,
  author    = {Bu{\c{c}}inca, Zana and Malaya, Maja Barbara and Gajos, Krzysztof Z.},
  title     = {To Trust or to Think: Cognitive Forcing Functions Can Reduce Overreliance on {AI} in {AI}-assisted Decision-making},
  booktitle = {Proceedings of the ACM on Human-Computer Interaction},
  year      = {2021},
  publisher = {Association for Computing Machinery},
  address   = {New York, NY, USA},
  doi       = {10.1145/3449287}
}

@inproceedings{park2023generative,
  author    = {Park, Joon Sung and O'Brien, Joseph C. and Cai, Carrie J. and Morris, Meredith Ringel and Liang, Percy and Bernstein, Michael S.},
  title     = {Generative Agents: Interactive Simulacra of Human Behavior},
  booktitle = {Proceedings of the 36th Annual ACM Symposium on User Interface Software and Technology},
  year      = {2023},
  publisher = {Association for Computing Machinery},
  address   = {New York, NY, USA},
  doi       = {10.1145/3586183.3606763}
}

@inproceedings{shaikh2025creatinggeneralusermodels,
  author    = {Shaikh, Omar and Sapkota, Shardul and Rizvi, Shan and Horvitz, Eric and Park, Joon Sung and Yang, Diyi and Bernstein, Michael S.},
  title     = {Creating General User Models from Computer Use},
  booktitle = {Proceedings of the 38th Annual ACM Symposium on User Interface Software and Technology},
  year      = {2025},
  publisher = {Association for Computing Machinery},
  address   = {New York, NY, USA},
  doi       = {10.1145/3746059.3747722}
}

@inproceedings{cheng2025conversational,
  author    = {Cheng, Yi Fei and Shirado, Hirokazu and Kasahara, Shunichi},
  title     = {Conversational Agents on Your Behalf: Opportunities and Challenges of Shared Autonomy in Voice Communication for Multitasking},
  booktitle = {Proceedings of the 2025 CHI Conference on Human Factors in Computing Systems},
  year      = {2025},
  publisher = {Association for Computing Machinery},
  address   = {New York, NY, USA},
  doi       = {10.1145/3706598.3714017}
}

@misc{gani2025klarna,
  author       = {Gani, Aisha S.},
  title        = {When Customers Dial {Klarna}'s Hotline, An {AI} {CEO} Picks Up},
  howpublished = {Bloomberg Tech In Depth (newsletter)},
  year         = {2025},
  month        = sep,
  day          = {10},
  url          = {https://www.bloomberg.com/news/newsletters/2025-09-10/when-customers-dial-klarna-s-hotline-an-ai-ceo-picks-up}
}

@misc{simile2026,
  author       = {{The Simile Team}},
  title        = {The Simulation Company},
  howpublished = {Simile Blog},
  year         = {2026},
  month        = feb,
  day          = {12},
  url          = {https://simile.ai/blog/the-simulation-company}
}

@misc{park2024generative,
  author    = {Park, Joon Sung and Zou, Carolyn Q. and Shaw, Aaron and Hill, Benjamin Mako and Cai, Carrie and Morris, Meredith Ringel and Willer, Robb and Liang, Percy and Bernstein, Michael S.},
  title     = {Generative Agent Simulations of 1,000 People},
  year      = {2024},
  doi       = {10.48550/arXiv.2411.10109}
}

@inproceedings{hu2026boss,
  author    = {Hu, Qing and Xiao, Qing and Cao, Hancheng and Shen, Hong},
  title     = {When Your Boss Is an {AI} Bot: Exploring Opportunities and Risks of Manager Clone Agents in the Future Workplace},
  booktitle = {Proceedings of the 2026 CHI Conference on Human Factors in Computing Systems},
  year      = {2026},
  publisher = {Association for Computing Machinery},
  address   = {New York, NY, USA},
  doi       = {10.48550/arXiv.2509.10993},

}

\end{document}